\documentclass[aps,prb,twocolumn]{revtex4}
\usepackage{amsmath}
\usepackage{graphicx}
\begin{document}

\title{Spin-Motive Forces and Current-Induced Torques in Ferromagnets}

\author{Kjetil M. D. Hals$^1$ and Arne Brataas$^2$}
\affiliation{$^1$ Niels Bohr International Academy and the Center for Quantum Devices, Niels Bohr Institute, University of Copenhagen, 2100 Copenhagen, Denmark. \\
$^2$ Department of Physics, Norwegian University of Science and  Technology, NO-7491, Trondheim, Norway. }

\begin{abstract}
In metallic ferromagnets, the spin-transfer torque and spin-motive force are known to exhibit a reciprocal relationship. Recent experiments on ferromagnets with strong spin-orbit coupling have revealed a rich complexity in the interaction between itinerant charge carriers and magnetization, but a full understanding of this coupled dynamics is lacking. Here, we develop a general phenomenology of the two reciprocal processes of charge pumping by spin-motive forces and current-driven magnetization dynamics. The formalism is valid for spin-orbit coupling of any strength and presents a systematic scheme for deriving all possible torque and charge-pumping terms that obey the symmetry requirements imposed by the point group of the system. We demonstrate how the different charge pumping and torque contributions are connected via the Onsager reciprocal relations. The formalism is applied to two important classes of systems: isotropic ferromagnets with non-uniform magnetization and homogeneous ferromagnets described by the point group $C_{2v}$.     
\end{abstract}
\maketitle

\section{Introduction}
A current of electrons that enters a ferromagnet can produce a torque on the magnetization due to the transfer of spin angular momentum from the itinerant carriers to the magnetic system. Following the first theoretical predictions reported in 1996 by Berger and Slonczewski,~\cite{Berger, Slon} the spin-transfer torque (STT) has garnered abundant experimental evidence.~\cite{Ralph:jmmm08} Today, current-driven magnetization dynamics is a highly active research area with a large application potential in magnetoelectronic devices (e.g., the first STT-RAM entered the marketplace in 2012). ~\cite{Ralph:jmmm08} 

The current-driven dynamics of metallic ferromagnets is described by the Laudau-Lifshitz-Gilbert-Slonczewski (LLGS) equation:~\cite{Ralph:jmmm08} 
\begin{equation}
\dot{\mathbf{m}} = -\gamma \mathbf{m}  \times \mathbf{H}_{\rm eff} + \mathbf{m}  \times \boldsymbol{\alpha} \dot{ \mathbf{m}}  + \boldsymbol{\tau} . \label{Eq:LLG}
\end{equation}
Here, $\mathbf{m} (\mathbf{r}, t)$ is a unit vector along the magnetization $\mathbf{M} (\mathbf{r}, t)= M_s \mathbf{m} (\mathbf{r}, t)$, and  $\gamma$ is the gyromagnetic ratio. The first term on the right side of Eq.~\eqref{Eq:LLG} is the torque from the effective field $\mathbf{H}_{\rm eff}(\mathbf{r}, t)= -(1/M_s)\delta F \left[ \mathbf{m}\right] / \delta\mathbf{m}(\mathbf{r}, t)$, which is given by the magnetic free energy $ F \left[ \mathbf{m}\right]$. The second term, which is parameterized by the second-rank Gilbert damping tensor $\boldsymbol{\alpha}$, describes the magnetization dissipation, whereas the last term represents the current-induced torque. 

\begin{figure}[t] 
\centering 
\includegraphics[scale=1.0]{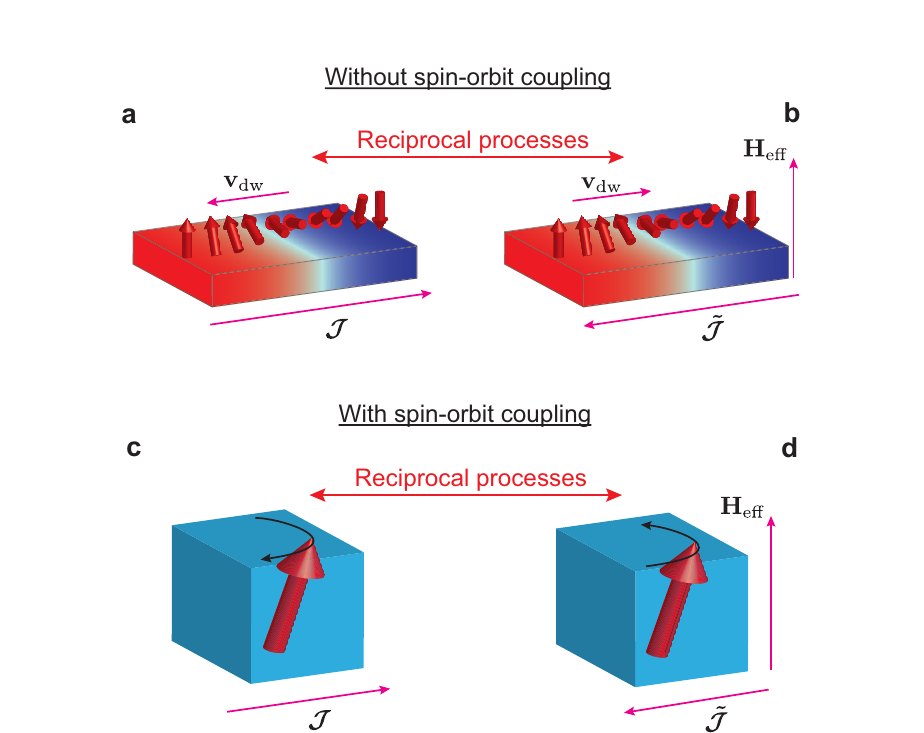}  
\caption{(Color online). (a)-(b) In metallic ferromagnets with negligible SOC, the magnetic texture is required to have a coupling between the itinerant quasi-particles and the magnetization.
(a) The current-driven domain wall motion  is reciprocal to (b) the spin-motive force induced by a moving domain wall. (c)-(d) In ferromagnets with broken spatial inversion symmetry, SOC mediates a coupling between the magnetization
and the itinerant quasi-particles, even without a magnetic textures.  (c) The SOT on a uniform magnetization is reciprocal to (d) charge pumping by homogeneous magnetization precession.   }
\label{Fig1} 
\end{figure} 

In systems with negligible intrinsic spin-orbit coupling (SOC),\cite{comment1} the current-induced torque is to lowest order in the magnetization gradients and to linear order in the applied current density $\boldsymbol{\mathcal{J}}$ given by~\cite{Ralph:jmmm08}
\begin{equation}
\boldsymbol{\tau}^{\rm STT} =  - \left( 1 - \beta\mathbf{m}\times   \right)  \left( \mathbf{v}_s\cdot\boldsymbol{\nabla} \right) \mathbf{m}, \label{Eq:STT}
\end{equation}
where $\mathbf{v}_s$ is proportional to $\boldsymbol{\mathcal{J}}$ and the spin polarization of the current. Importantly, we note that no effect of the crystal symmetry is reflected in Eq.~\eqref{Eq:STT}. The torque is fully rotationally symmetric under \emph{separate rotations} of the magnetization (the spin space) and spatial coordinates (the real space). Furthermore, the torque vanishes in a homogeneous ferromagnet because the gradients of the magnetization are zero in this case. The first part of Eq.~\eqref{Eq:STT} represents a reactive STT because it preserves the time reversal symmetry of the LLGS equation~\eqref{Eq:LLG}. The second term, which is parameterized by $\beta$, breaks the time reversal symmetry of Eq.~\eqref{Eq:LLG} and therefore describes a dissipative STT. Its magnitude is governed by the spin-flip scattering rate induced by extrinsic SOC and magnetic impurities.

Recent experimental~\cite{Chernyshov:nature09, Miron:nature10, Miron:n2011, Miron:nm2011, Fang:nn2013, Liu:prl2012, Liu:science2012,  Emori:arxiv2013, Thomas:nnt2013, Garello:nn2013, Haazen:nm2013, Fan:nc2013, Kurebayashi:arxiv2013, Emori:prb2014} and theoretical works~\cite{Bernevig:prb05, Manchon:prb08, Obata:prb08, Garate:prb09, Hals:epl10, Pesin:prb2012, Bijl:prb2012, Wang:prl2012, Freimuth:prb2014} have shown that the torque in Eq.~\eqref{Eq:STT} does not describe the current-driven dynamics of metallic ferromagnets with large intrinsic SOC. In asymmetric systems, the SOC combined with an external electric field produces an out-of-equilibrium spin density, which yields a magnetization torque via the exchange interaction. This torque, which originates solely from the SOC,  is referred to as a spin-orbit torques (SOTs). Unlike the conventional STT in Eq.~\eqref{Eq:STT}, SOTs can induce magnetization dynamics even in the absence of magnetization gradients, and their particular forms are determined by the crystal symmetry. In Ref.~\onlinecite{Hals:prb2013b}, a general phenomenology of current-induced torques in systems with arbitrarily strong SOC was developed. Based on NeumannÕs principle,\cite{Birss:book} which is a fundamental symmetry principle of crystal physics, the phenomenological theory outlines a systematic scheme to derive all possible torque terms that are allowed by symmetry. The formalism shows that the SOC introduces several new torques in addition to the reactive and dissipative STTs in Eq.~\eqref{Eq:STT}. Additionally, the reactive and dissipative STTs become renormalized by the SOC, and the simplest extension of Eq.~\eqref{Eq:STT} to account for the effects of the SOC is~\cite{Hals:prb2013b} 
\begin{equation}
\boldsymbol{\tau}= -\gamma\mathbf{m}\times \mathbf{H}_c ,\label{Eq:TorqueExtended1} 
\end{equation}
where the current-induced field $\mathbf{H}_c = \boldsymbol{\eta}\boldsymbol{\mathcal{J}}$ is determined by the second-rank tensor $\boldsymbol{\eta}= \left[  \eta_{ij} \right]$, which is given by 
\begin{equation}
\eta_{ij}=  \Lambda^{(r)}_{ij} + \Lambda^{(d)}_{ijk} m_k + \beta_{ijkl}\partial_k m_l + P_{ijkln}m_k \partial_l m_n . \label{Eq:TorqueExtended2}
\end{equation}
Here, the tensorial forms of $\Lambda^{(r)}_{ij}$, $\Lambda^{(d)}_{ijk}$, $\beta_{ijkl}$, and $P_{ijkln}$ are governed by the crystal symmetry of the system.
The tensors $\Lambda^{(r)}_{ij}$ and $\Lambda^{(d)}_{ijk}$ describe the reactive and dissipative SOTs, respectively, which are present only in systems with broken spatial inversion symmetry, whereas the two tensors $P_{ijkln}$ and $\beta_{ijkl}$ represent generalizations of the reactive and dissipative STTs in Eq.~\eqref{Eq:STT}. Here and below, a summation of repeated indices is implied.
 
Reciprocal processes are well known in several fields of physics. The most familiar example may be the thermoelectric effect, where a thermal gradient induces an electric current, or vice versa, an electric field produces a flow of heat. The linear response coefficients of reciprocal processes are connected via the Onsager reciprocal relations.\cite{Groot:book} Thus, by 
measuring one of these two processes, one also gains information about the strength and symmetry of the reciprocal phenomenon.

In the reciprocal effect of the current-induced torque, charge currents are induced by time variations in the magnetization, and the forces that induce such currents are referred to as spin-motive forces (Fig.~\ref{Fig1}). 
In the absence of a large intrinsic SOC and to the lowest order in the magnetization precession frequency and magnetization gradients, the pumped charge current is~\cite{Volovik, Barnes:prl07, Tatara:prl07, Tserkovnyak:prb08, Duine:prb08, Yang:prl09}
\begin{equation}
\tilde{\mathcal{J}}_i = \tilde{P}\mathbf{m}\cdot \left( \partial_t \mathbf{m}\times \partial_i \mathbf{m} \right) + \tilde{\beta} \partial_t \mathbf{m}\cdot \partial_i \mathbf{m} . \label{Eq:ChargePumpingSTT}
\end{equation}
Eq.~\eqref{Eq:ChargePumpingSTT} is reciprocal to the STT in Eq.~\eqref{Eq:STT} (Fig.~\ref{Fig1}a-b). The first term, which is parameterized by $\tilde{P}$, is the reciprocal process of the reactive STT, whereas the second term $\sim\tilde{\beta}$ is reciprocal to the dissipative STT. The charge-pumping effect of Eq.~\eqref{Eq:ChargePumpingSTT} has been experimentally detected in a ferromagnetic nanostrip, where an electric signal induced by moving domain walls was observed.\cite{Yang:prl09} 

The general form of the current-induced torque in Eqs.~\eqref{Eq:TorqueExtended1}-\eqref{Eq:TorqueExtended2} indicates that strong intrinsic SOC will also increased the complexity in charge pumping. For example, the SOTs proportional to $\Lambda^{(r)}_{ij}$ and $\Lambda^{(d)}_{ijk}$ imply via the reciprocity relations that a homogeneous precessing magnetization pumps a charge current in asymmetric ferromagnets (Fig.~\ref{Fig1}c-d). This effect, which arises solely from the SOC, was recently observed in strained (Ga,Mn)As~\cite{Chiara:nn2015} and has been studied in several theoretical works.~\cite{Hals:epl10, Kim:prl2012, Hals:prb2013a, Tatara:prb2013, Freimuth:arxiv2014} However, the effects of the SOC on the charge pumping in ferromagnets with an arbitrary magnetic texture (e.g., magnetic domain walls, skyrmions or vortices) and the relationship between different charge pumping contributions and the current-induced torques remain unknown.  

In this paper, we extend the theory of Ref.~\onlinecite{Hals:prb2013b} and develop a general phenomenology of the coupled dynamics between itinerant quasi-particles and magnetization in metallic ferromagnets. The formalism is valid for intrinsic SOC of any strength. Our phenomenology is based on Neumann's principle, which is used to derive the possible forms of charge pumping and current-induced torques allowed by the symmetry of the system.
The Onsager reciprocal relations are then used to link the different charge pumping and torque terms. We apply the formalism to two classes of systems: {\it i)} isotropic ferromagnets with non-uniform magnetization and {\it ii)}  
homogeneous ferromagnets with Rashba and Dresselhaus SOC. We show that the conventional LLGS phenomenology (i.e., Eqs.~\eqref{Eq:STT} and \eqref{Eq:ChargePumpingSTT}) is obtained as the non-relativistic limit for isotropic systems.  In these systems, the SOC introduces two new dissipative torques/pumping terms and four new reactive torques/pumping terms in addition to the conventional terms: to the best of our knowledge, several of these additional terms have not been previously reported. Importantly, these new terms lead to different dynamics for Bloch and Neel domain walls. For Rashba and Dresselhaus SOC, we derive the reactive and dissipative SOTs and charge pumping, and we show that the results are consistent with recent experimental observations in strained (Ga,Mn)As epilayers.

This paper is organized as follows. First, we introduce Neumann's principle and the Onsager reciprocal relations in Sec.~\ref{Sec:GeneralTheory} before deriving the general phenomenology 
of current-induced torques and charge pumping in Sec.~\ref{Sec:SymRel}. The formalism is applied to isotropic and asymmetric ferromagnets in Sec.~\ref{Sec:Iso} and Sec.~\ref{Sec:C2v}.
A summary is provided in Sec.~\ref{Sec:Summary}.

\section{Fundamental Symmetry Principles} \label{Sec:GeneralTheory}
Our phenomenology is based on two fundamental symmetry principles in condensed matter physics: Neumann's principle and Onsager's reciprocity relations.
The present section gives a brief introduction to the general theory of these principles.  
We refer the reader to Refs.~\onlinecite{Birss:book} and \onlinecite{Groot:book} for a more comprehensive treatment.

\subsection{Transformation properties of tensors}
In the following discussion, it is convenient to introduce some basic concepts regarding tensors and vectors.
  
Let $\{\boldsymbol{\mathcal{R}} \in \mathcal{G}\}$ denote a set of $3\times 3$ matrices, which represent the crystallographic point group $\mathcal{G}$. 
The symmetry operation $\boldsymbol{\mathcal{R}}$ can be either be a proper rotation with a determinant $|\mathcal{R}|=1$ or an improper rotation, which changes the orientation of the coordinate
system and therefore has a determinant $|\mathcal{R}|= -1$. 

Vectors can be classified into two groups according to how they transform under a 
coordinate transformation: $\acute{\mathbf{r}}=\boldsymbol{\mathcal{R}}\mathbf{r}$. A polar vector $\mathbf{u}$ transforms as $\acute{\mathbf{u}}= \boldsymbol{\mathcal{R}}\mathbf{u}$, whereas a pseudo vector transforms as $\acute{\mathbf{u}}= |\mathcal{R}| \boldsymbol{\mathcal{R}}\mathbf{u}$. 
Importantly, polar vectors change sign under the spatial inversion operation $\mathcal{R}_{ij}= -\delta_{ij}$ ($\delta_{ij}$: the Kronecker delta), whereas pseudo vectors remain unchanged.
Examples of pseudo and polar vectors are the magnetization and charge current, respectively, both of which are central to our discussion.  
 
Vectors are tensors of rank one. 
The classification of vectors according to their transformation properties can be generalized to tensors of any rank.
A polar tensor $T_{ij\ldots k}$ transforms as
\begin{equation} 
T_{ij\ldots k}= \mathcal{R}_{i\alpha} \mathcal{R}_{j\beta}\ldots \mathcal{R}_{k\gamma}T_{\alpha\beta\ldots \gamma},  \label{Eq:PolarTensor}
\end{equation}
under a coordinate transformation $\acute{\mathbf{r}}=\boldsymbol{\mathcal{R}}\mathbf{r}$, whereas an axial tensor (or pseudo tensor) transforms as 
\begin{equation}
T_{ij\ldots k}= |\mathcal{R}| \mathcal{R}_{i\alpha} \mathcal{R}_{j\beta}\ldots \mathcal{R}_{k\gamma}T_{\alpha\beta\ldots \gamma}. \label{Eq:AxialTensor}
\end{equation}

Next, we introduce the concepts of thermodynamic forces and fluxes, which are required in the formulation of Onsager's reciprocity relations.

\subsection{Thermodynamic forces and fluxes}
Consider a system described by a set of thermodynamic variables $\{ q_i\}$. Examples of such variables include the charge density, spin density, and magnetization.
Let $f_i$ denote the thermodynamic force that induces a flux $J_i$ in the quantity $q_i$. In general, the fluxes and forces are not uniquely defined,  but for a system with uniform temperature, they should be properly normalized to yield the entropy generation
$\dot{S}$ via the relation
\begin{equation}
T\dot{S} = J_i f_i. \label{Eq:EntropyGeneration}
\end{equation}

In the present work, we consider the linear response regime and treat the dynamics in the local approximation, where all linear response coefficients are
assumed to be local in space and time, i.e., the flux $\mathbf{J} (\mathbf{r}, t)$ depends only on the value of the force $\mathbf{f} (\mathbf{r}, t)$ at the space-time position $(\mathbf{r}, t)$.  
Then, the fluxes are related to the forces by the equations
 \begin{equation}
 J_i (\mathbf{r}, t) = L_{ij} (\mathbf{r})  f_j (\mathbf{r}, t) , \label{Eq:LinearResponse}
 \end{equation}
 where the linear response coefficients $L_{ij}$ depend on the equilibrium value of the variables $\{ q_i\}$. 
 The local approximation implies that the magnetization slowly varies over typical transport length scales such as the mean free path,
 spin-flip diffusion length, and transverse-spin decoherence length.

\subsection{Neumann's principle}
The crystal symmetry of the system reduces the number of independent coefficients $L_{ij}$. 
Formally, this reduction is expressed by Neumann's principle, which states that \emph{ ''any type of symmetry that is exhibited by the point group of the crystal is possessed by every physical property of the crystal''. }
Neumann's principle implies that the response coefficients $L_{ij}$ satisfy the symmetry relations    
\begin{equation}
L_{ij}(\acute{\mathbf{H}},\acute{\mathbf{M}}) =  |\mathcal{R}|^{\zeta_i + \zeta_j}     \mathcal{R}_{i \alpha} \mathcal{R}_{j \beta}  L_{\alpha \beta }(\mathbf{H},\mathbf{M})   , \label{Eq:Neumann}
\end{equation}
where $\zeta_i= 0$ ($\zeta_i= 1$) if $q_i$ is a component of a vector (a pseudo vector) and $\mathbf{H}$ represents an external magnetic field.

\subsection{Onsager reciprocal relations}
The off-diagonal terms of the response matrix $\mathbf{L}= [L_{ij}]$ describe how a thermodynamic force $f_j$ of a variable $q_j$ induces a response in another thermodynamic variable $q_i$.
The reciprocal phenomenon of this effect is the process by which the thermodynamic force $f_i$ induces a response in the quantity $q_j$. 
The linear response coefficients of the reciprocal phenomena are connected via the Onsager reciprocal relations, which imply that
\begin{equation}
L_{ij}(\mathbf{H},\mathbf{M})= \epsilon_i\epsilon_jL_{ji}(-\mathbf{H},-\mathbf{M}) . \label{Eq:Onsager1}
 \end{equation}
Here,  $\epsilon_i=1$ ($\epsilon_i=-1$) if $q_i$ is even (odd) under time reversal. 

The symmetry relationships implied by Eqs.~\eqref{Eq:Neumann} and \eqref{Eq:Onsager1}  considerably simplify the possible allowed forms of the response matrix $\mathbf{L}$ and the number of
independent matrix elements.

\section{Symmetry Relations of Charge Pumping and Current-Induced Torques} \label{Sec:SymRel}
We now use Neumann's principle to derive the possible expressions for the charge pumping and the current-induced torque. 
Then, the Onsager reciprocal relations are used to derive relationships between these expressions.

To use the Onsager reciprocal relations, the LLGS equation \eqref{Eq:LLG} should be rewritten such that the time derivative of the magnetization only appears 
on the left-hand side of the equation. We refer to this as the Landau-Lifshitz (LL) form of the equation. To facilitate this reformulation, it is convenient to introduce the matrix
\begin{equation}
O_{ij}= \epsilon_{i k j} m_k,
\end{equation}    
where $\epsilon_{i k l }$ is the  Levi-Civita tensor.
Then, the LLG equation \eqref{Eq:LLG} in the LL form becomes 
\begin{equation}
\dot{\mathbf{m}} = -\gamma \left[  1 -  \mathbf{O} \boldsymbol{\alpha} \right]^{-1} \mathbf{O} \mathbf{H}_{\rm eff} + \boldsymbol{\tau}^{\rm LL}  , \label{Eq:LL}
\end{equation}
where the torque in the LL formulation is 
\begin{equation}
\boldsymbol{\tau}^{\rm LL} =  \left[  1 - \mathbf{O} \boldsymbol{\alpha} \right]^{-1} \boldsymbol{\tau} . 
\end{equation}

First, we determine the different fluxes and forces. We consider a metallic ferromagnet at constant temperature, where the magnetization is coupled to the itinerant charge carriers.
The magnetic system is described by the magnetization $\mathbf{m}$, whereas the charge system is characterized by the charge density $\rho$. The associated fluxes of these quantities are
the time derivative of the magnetization $\dot{\mathbf{m}}$ and the charge current density $\boldsymbol{\mathcal{J}}$, and the corresponding forces that induce the fluxes are the effective field $M_s \mathbf{H}_{\rm eff}$ and the electric field $\mathbf{E}$.
These fluxes and forces are summarized in Table~\ref{tab:1}, where the values of the parameters $\zeta_i$ and $\epsilon_i$ in Eqs.~\eqref{Eq:Neumann}-\eqref{Eq:Onsager1} are also provided.     
\begin{table}
\centering
\caption{Fluxes and forces in metallic ferromagnets. }
\begin{tabular}{ l c c }
\hline\hline \\
 & \ \ \ \ \ \ Magnetic System\ \ \ \ \ \ &  \ \ \ \ \ \ Charge System \ \ \ \ \ \  \\
 \hline \\
 Flux & $\dot{\mathbf{m}}$ & $\boldsymbol{\mathcal{J}}$ \\ 
 Force & $\mathbf{f}_m= M_s \mathbf{H}_{\rm eff}$ & $\mathbf{E}$ \\
 $\zeta_i$  &  1 & 0 \\
 $\epsilon_i$  & -1 & 1\\ 
 \hline \hline	 
\end{tabular}
\label{tab:1}
\end{table}
In the linear response regime, these fluxes and forces are related via 
\begin{equation}
\begin{pmatrix}
\dot{\mathbf{m}} \\
\boldsymbol{\mathcal{J}} 
\end{pmatrix} 
= 
\begin{pmatrix}
\mathbf{L}^{mm} & \mathbf{L}^{m\rho}  \\
\mathbf{L}^{\rho m} & \mathbf{L}^{\rho \rho}
\end{pmatrix}
\begin{pmatrix}
\mathbf{f}_m \\
\mathbf{E}
\end{pmatrix} .
\end{equation}
The $3\times 3$ matrix $\mathbf{L}^{mm}$ describes the magnetization dynamics driven by the effective field and is determined by  Eq.~\eqref{Eq:LL}:
 \begin{equation}
\mathbf{L}^{mm} \left(   \mathbf{m}, \boldsymbol{\nabla} \mathbf{m} \right) =  -\frac{\gamma}{M_s} \left[  1 -  \mathbf{O} \boldsymbol{\alpha} \right]^{-1} \mathbf{O} . 
\end{equation}
The $3\times 3$ matrix $ \mathbf{L}^{\rho \rho}$ is the conductivity tensor 
 \begin{equation}
\mathbf{L}^{\rho \rho} \left(   \mathbf{m}, \boldsymbol{\nabla} \mathbf{m} \right) = \boldsymbol{\sigma} \left(   \mathbf{m}, \boldsymbol{\nabla} \mathbf{m} \right) ,
 \end{equation}
whereas the $3\times 3$ matrices $\mathbf{L}^{\rho m }$ and $\mathbf{L}^{m \rho}$ describe the charge pumping and current-induced torque, respectively. 

Based on the general theory presented in Sec.~\ref{Sec:GeneralTheory},
we derive the symmetry properties of these four submatrices. 
Under a symmetry transformation $\boldsymbol{\mathcal{R}}\in \mathcal{G}$, Neumann's principle implies that
\begin{eqnarray}
\mathbf{L}^{mm} \left(  \acute{\mathbf{m}}, \acute{\boldsymbol{\nabla}} \acute{\mathbf{m} } \right) &=&  \boldsymbol{\mathcal{R}} \mathbf{L}^{mm} \left(   \mathbf{m}, \boldsymbol{\nabla} \mathbf{m} \right) \boldsymbol{\mathcal{R}}^T,\label{Eq:L1}  \\
\mathbf{L}^{\rho\rho} \left(  \acute{\mathbf{m}}, \acute{\boldsymbol{\nabla}} \acute{\mathbf{m} } \right) &=&  \boldsymbol{\mathcal{R}} \mathbf{L}^{\rho\rho} \left(   \mathbf{m}, \boldsymbol{\nabla} \mathbf{m} \right) \boldsymbol{\mathcal{R}}^T,\label{Eq:L2}  \\
\mathbf{L}^{\rho m} \left(  \acute{\mathbf{m}}, \acute{\boldsymbol{\nabla}} \acute{\mathbf{m} } \right) &=& |\mathcal{R}| \boldsymbol{\mathcal{R}} \mathbf{L}^{\rho m} \left(   \mathbf{m}, \boldsymbol{\nabla} \mathbf{m} \right) \boldsymbol{\mathcal{R}}^T,\label{Eq:L3} \\
\mathbf{L}^{m\rho} \left(  \acute{\mathbf{m}}, \acute{\boldsymbol{\nabla}} \acute{\mathbf{m} } \right) &=&  |\mathcal{R}| \boldsymbol{\mathcal{R}} \mathbf{L}^{m \rho} \left(   \mathbf{m}, \boldsymbol{\nabla} \mathbf{m} \right) \boldsymbol{\mathcal{R}}^T, \label{Eq:L4}
\end{eqnarray} 
whereas the Onsager reciprocal relations yield the following relationships for the off-diagonal matrix elements
\begin{eqnarray}
L_{ij}^{m m} \left(   \mathbf{m}, \boldsymbol{\nabla} \mathbf{m} \right)  &=& L_{ji}^{m m} \left(   -\mathbf{m}, -\boldsymbol{\nabla} \mathbf{m} \right) ,\label{Eq:OnsagerRel2} \\
L_{ij}^{\rho \rho} \left(   \mathbf{m}, \boldsymbol{\nabla} \mathbf{m} \right)  &=&  L_{ji}^{\rho \rho} \left(   -\mathbf{m}, -\boldsymbol{\nabla} \mathbf{m} \right) .\label{Eq:OnsagerRel3}\\
L_{ij}^{m\rho} \left(   \mathbf{m}, \boldsymbol{\nabla} \mathbf{m} \right)  &=& -  L_{ji}^{m\rho} \left(   -\mathbf{m}, -\boldsymbol{\nabla} \mathbf{m} \right) .\label{Eq:OnsagerRel1} 
\end{eqnarray} 

The current-induced torque and charge pumping are usually expressed in terms of the applied charge current density and the time derivative of the magnetization. 
To linear order in the magnetization precession frequency and the out-of-equilibrium charge current density, the most general forms for the charge pumping and torque in the local approximation are
\begin{eqnarray}
\boldsymbol{\mathcal{\tilde{J}}} &=& \boldsymbol{\tilde{\eta}}\dot{\mathbf{m}} , \label{Eq:J1} \\
\boldsymbol{\tau} &=& -\gamma \mathbf{m}\times \mathbf{H}_c, \nonumber \\ 
\mathbf{H}_c &=& \boldsymbol{\eta} \boldsymbol{\mathcal{J}} . \label{Eq:M1} 
\end{eqnarray}
Here, $\boldsymbol{\tilde{\eta}}= \left[  \tilde{\eta}_{ij} \right]$ and $\boldsymbol{\eta}= \left[  \eta_{ij} \right]$ are second-rank tensors that act to the right on $\dot{\mathbf{m}}$ and $\boldsymbol{\mathcal{J}}$, respectively.
The tensors depend on the equilibrium value of the magnetization and its gradients, i.e., 
$\boldsymbol{\tilde{\eta}}= \boldsymbol{\tilde{\eta}} \left( \mathbf{m}, \boldsymbol{\nabla} \mathbf{m} \right) $  and $\boldsymbol{\eta}= \boldsymbol{\eta} \left( \mathbf{m}, \boldsymbol{\nabla} \mathbf{m} \right) $. 
$\mathbf{H}_c$ is the effective field produced by the current density and contains both dissipative and reactive contributions to the torque.
Note that Eq.~\eqref{Eq:M1} refers to the current-induced torque in the LLGS equation \eqref{Eq:LLG}. In the LL formulation  \eqref{Eq:LL}, the torque becomes
\begin{equation}
\boldsymbol{\tau}^{\rm LL} =  -\gamma \left[  1 -  \mathbf{O} \boldsymbol{\alpha} \right]^{-1} \mathbf{O}\mathbf{H}_c= M_s \mathbf{L}^{mm} \boldsymbol{\eta} \boldsymbol{\mathcal{J}} .  \label{Eq:M1b} 
\end{equation}  

To derive the symmetry properties of the two tensors $\boldsymbol{\tilde{\eta}}$ and $\boldsymbol{\eta}$, we consider an out-of-equilibrium current density caused by an external electric field $\mathbf{E}$ and magnetization dynamics driven by an effective magnetic field $\mathbf{H}_{\rm eff}$.
Using $\boldsymbol{\mathcal{J}}= \mathbf{L}^{\rho\rho} \mathbf{E}$ and $\dot{\mathbf{m}} = \mathbf{L}^{m m} \mathbf{f}_m$ in Eqs.~\eqref{Eq:J1} and \eqref{Eq:M1b}, we find that  $\boldsymbol{\tilde{\eta}}$ and $\boldsymbol{\eta}$ are related to the response matrices 
$\mathbf{L}^{\rho m}$ and $\mathbf{L}^{m \rho}$ by the equations  
\begin{eqnarray}
\mathbf{L}^{\rho m} &=& \boldsymbol{\tilde{\eta}}\mathbf{L}^{mm} , \label{Eq:LambdaL3} \\
\mathbf{L}^{m \rho} &=& M_s\mathbf{L}^{mm}  \boldsymbol{\eta}\mathbf{L}^{\rho\rho} .\label{Eq:etaL4}
\end{eqnarray} 
In Eqs.~\eqref{Eq:LambdaL3}-\eqref{Eq:etaL4}, the matrices  $\mathbf{L}^{mm}$ and
$\mathbf{L}^{\rho\rho}$ transform according to  Eqs.~\eqref{Eq:L1}-\eqref{Eq:L2} under a symmetry transformation $\boldsymbol{\mathcal{R}}$. 
Thus, for $\mathbf{L}^{\rho m}$ and $\mathbf{L}^{m \rho}$ to satisfy the symmetry relationships implied by Eqs.~\eqref{Eq:L3} and \eqref{Eq:L4}, the tensors 
$\boldsymbol{\tilde{\eta}}$ and $\boldsymbol{\eta}$ must transform as
\begin{eqnarray}
\boldsymbol{\tilde{\eta}} \left(  \acute{\mathbf{m}}, \acute{\boldsymbol{\nabla}} \acute{\mathbf{m} } \right) &=& |\mathcal{R}|  \boldsymbol{\mathcal{R}} \boldsymbol{\tilde{\eta}} \left(   \mathbf{m}, \boldsymbol{\nabla} \mathbf{m} \right) \boldsymbol{\mathcal{R}}^T,\label{Eq:SymLambda}  \\
\boldsymbol{\eta} \left(  \acute{\mathbf{m}}, \acute{\boldsymbol{\nabla}} \acute{\mathbf{m} } \right) &=& |\mathcal{R}|  \boldsymbol{\mathcal{R}} \boldsymbol{\eta} \left(   \mathbf{m}, \boldsymbol{\nabla} \mathbf{m} \right) \boldsymbol{\mathcal{R}}^T.\label{Eq:SymEta} 
\end{eqnarray}
Eqs.~\eqref{Eq:SymLambda}-\eqref{Eq:SymEta} determine the tensorial forms of $\boldsymbol{\tilde{\eta}}$ and $\boldsymbol{\eta}$.  
 
In addition to the symmetry requirements imposed by Neumann's principle, the Onsager reciprocal relations in Eq.~\eqref{Eq:OnsagerRel1} yield the equation 
\begin{widetext}
\begin{equation}
M_s\mathbf{L}^{mm}\left(   \mathbf{m}, \boldsymbol{\nabla} \mathbf{m} \right)  \boldsymbol{\eta} \left(   \mathbf{m}, \boldsymbol{\nabla} \mathbf{m} \right)  \mathbf{L}^{\rho\rho} \left(   \mathbf{m}, \boldsymbol{\nabla} \mathbf{m} \right)  =  -\mathbf{L}^{mm} \left(   -\mathbf{m}, -\boldsymbol{\nabla} \mathbf{m} \right)^T \boldsymbol{\tilde{\eta}} \left( -  \mathbf{m}, - \boldsymbol{\nabla} \mathbf{m} \right)^T  . \label{Eq:ResRel1} 
\end{equation} 
\end{widetext} 
Using Eq.~\eqref{Eq:OnsagerRel2}, we identify the following reciprocal relationship between $\boldsymbol{\tilde{\eta}} $ and $\boldsymbol{\eta}$:
\begin{equation}
M_s \boldsymbol{\eta} \left(   \mathbf{m}, \boldsymbol{\nabla} \mathbf{m} \right)  \boldsymbol{\sigma} \left(   \mathbf{m}, \boldsymbol{\nabla} \mathbf{m} \right)  =  -\boldsymbol{\tilde{\eta}} \left( -  \mathbf{m}, - \boldsymbol{\nabla} \mathbf{m} \right)^T .  \label{Eq:ResRel2} 
\end{equation}
Here, we use the conventional expression $\boldsymbol{\sigma}$ for the conductivity tensor instead of $\mathbf{L}^{\rho\rho}$. 

Because the tensor $M_s\boldsymbol{\eta}\boldsymbol{\sigma}$ is directly connected to the charge pumping tensor $\boldsymbol{\tilde{\eta}} $ via the reciprocity relations, it is useful to introduce the notation
\begin{equation}
\boldsymbol{\hat{\eta}} \left(   \mathbf{m}, \boldsymbol{\nabla} \mathbf{m} \right)  \equiv M_s  \boldsymbol{\eta} \left(   \mathbf{m}, \boldsymbol{\nabla} \mathbf{m} \right)  \boldsymbol{\sigma} \left(   \mathbf{m}, \boldsymbol{\nabla} \mathbf{m} \right)  .  \label{Eq:EtaHat} 
\end{equation}
Eq.~\eqref{Eq:M1} implies that $\boldsymbol{\hat{\eta}}$ describes the effective field induced by the external electric field 
\begin{equation}
\mathbf{H}_c = \frac{1}{M_s}\boldsymbol{\hat{\eta}} \mathbf{E} . 
\end{equation}
The direction of the applied electric field is typically known in an experiments, whereas the local current density is determined by local magnetoresistance effects. Therefore, it is often
preferable in systems with large SOC to express the current-induced torque in terms of the electric field (using the tensor $\boldsymbol{\hat{\eta}}$).  

Using the transformation properties of the conductivity tensor and $\boldsymbol{\eta}$, one finds that $\boldsymbol{\hat{\eta}}$ satisfies the transformation rule
\begin{equation}
\boldsymbol{\hat{\eta}} \left(  \acute{\mathbf{m}}, \acute{\boldsymbol{\nabla}} \acute{\mathbf{m} } \right) = |\mathcal{R}|  \boldsymbol{\mathcal{R}} \boldsymbol{\hat{\eta}} \left(   \mathbf{m}, \boldsymbol{\nabla} \mathbf{m} \right) \boldsymbol{\mathcal{R}}^T,\label{Eq:SymEtaHat}
\end{equation} 

Next, we will use the relations \eqref{Eq:SymLambda},  \eqref{Eq:SymEta}, and \eqref{Eq:SymEtaHat} with Eqs.~\eqref{Eq:J1}-\eqref{Eq:M1} to develop a systematic scheme to derive all possible torque and charge pumping terms.

\subsection{Phenomenological expansion} \label{Sec:PhenEx}
In general, several tensors satisfy the symmetry requirements imposed by Eqs.~\eqref{Eq:SymLambda}, \eqref{Eq:SymEta} and \eqref{Eq:SymEtaHat}.
This situation is analogous to the case of finding the magnetic free energy that describes the magnetocrystalline anisotropy energy.
In that case, one has a free energy functional that satisfies a set of symmetry relations $F \left[ \mathbf{m}\right]= F \left[  |\mathcal{R}|  \boldsymbol{\mathcal{R}} \mathbf{m}\right]$ generated by the point group $\{ \boldsymbol{\mathcal{R}}\in \mathcal{G}\}$.   
It is common to approximate the free energy by performing an analytic expansion $F \left[ \mathbf{m}\right] \approx K_{ij}^{(2)} m_i m_j + K_{ijkl}^{(4)} m_i m_j m_k m_l + ...$ in powers of $\mathbf{m}$, where each term in the expansion satisfies the correct symmetry. Usually, the first harmonics in this expansion capture the magnetocrystalline anisotropy observed in experiments. 
The series expansion rapidly converges because higher-order terms have a higher degree of anisotropy and thus tend to average out  
due to thermal fluctuations. Additionally, the higher-order terms are believed to be of a higher order in the SOC. 

With this motivation, we perform a similar analytic expansion of the tensors  $\boldsymbol{\tilde{\eta}}$, $\boldsymbol{\hat{\eta}}$ and $\boldsymbol{\eta}$:
\begin{eqnarray}
\tilde{\eta}_{ij} &=&  \tilde{\Lambda}^{(r)}_{ij} + \tilde{\Lambda}^{(d)}_{ijk} m_k + \tilde{\beta}_{ijkl}\partial_k m_l + \tilde{P}_{ijkln}m_k \partial_l m_n , \nonumber \\
\hat{\eta}_{ij} &=&  \hat{\Lambda}^{(r)}_{ij} + \hat{\Lambda}^{(d)}_{ijk} m_k + \hat{\beta}_{ijkl}\partial_k m_l + \hat{P}_{ijkln}m_k \partial_l m_n , \nonumber \\
\eta_{ij} &=&  \Lambda^{(r)}_{ij} + \Lambda^{(d)}_{ijk} m_k + \beta_{ijkl}\partial_k m_l + P_{ijkln}m_k \partial_l m_n . \nonumber \\
&& \label{Eq:ExpLambdaEta}
\end{eqnarray}
In Eq.~\eqref{Eq:ExpLambdaEta}, we keep only the first harmonics, which represent generalizations of the torque and charge pumping expressions in Eqs.~\eqref{Eq:STT} and \eqref{Eq:ChargePumpingSTT}.  
The expansion can be continued to arbitrary high orders in $\mathbf{m}$ to fit the  experimental observations for a particular system.
Such higher-order harmonics represent the current-induced torques and charge pumping with higher degrees of anisotropy.

The Onsager reciprocal relations \eqref{Eq:ResRel2} imply the following relationships among the tensors in Eq.~\eqref{Eq:ExpLambdaEta}
\begin{eqnarray}
\hat{\Lambda}^{(r)}_{i j} &=& -\tilde{\Lambda}^{(r)}_{ji} , \label{Eq:Onsager_1}\\
 \hat{\Lambda}^{(d)}_{ijk}  &=&  \tilde{\Lambda}^{(d)}_{jik} ,  \label{Eq:Onsager_2} \\
 \hat{\beta}_{ijkl}  &=& \tilde{\beta}_{jikl} \label{Eq:Onsager_3} , \\
\hat{P}_{ijkln}  &=& -\tilde{P}_{jikln} . \label{Eq:Onsager_4}
\end{eqnarray}
Eqs. \eqref{Eq:Onsager_1}-\eqref{Eq:Onsager_4} connect the reciprocal terms in the phenomenological expansions of the charge pumping and current-induced torque.
For example, $\tilde{\Lambda}^{(r)}_{ij}$ represents the charge pumping process that is reciprocal to the torque given by the tensor  $\hat{\Lambda}^{(r)}_{ij}$.

In general, the conductivity tensor also depends on the local direction of the 
magnetization and its gradients.  To capture this dependency, one can perform a similar phenomenological 
expansion of the conductivity in powers of $\mathbf{m}$. An expansion of the diagonal tensor elements describes the
anisotropic magnetoresistance (AMR) effect, whereas an expansion of the off-diagonal elements describes the anomalous Hall effect (AHE).
By relating terms of equal order in $\mathbf{m}$ in the two expressions $M_s \boldsymbol{\eta}\boldsymbol{\sigma}$ and $\boldsymbol{\hat{\eta}}$,
one can derive relationships among different tensor elements in the expansions of $\boldsymbol{\eta}$ and $\boldsymbol{\hat{\eta}}$.   

The forms of the tensors in Eq.~\eqref{Eq:ExpLambdaEta} are determined by Eqs.~\eqref{Eq:SymLambda}-\eqref{Eq:SymEta} and \eqref{Eq:SymEtaHat}.
Because the tensors $\boldsymbol{\tilde{\eta}}$, $\boldsymbol{\hat{\eta}}$ and $\boldsymbol{\eta}$ transform similarly under a coordinate transformation, the 
terms in their phenomenological expansions also satisfy identical symmetry relations. 
Therefore, it is sufficient to write down the relationships satisfied by the tensors in the expansion for the current-induced torque: 
\begin{eqnarray}
\Lambda^{(r)}_{ij} &=& |\mathcal{R}| \mathcal{R}_{i i^{'}} \mathcal{R}_{j j^{'}}  \Lambda^{(r)}_{i^{'} j^{'}} , \label{Eq:SymRel_1} \\
\Lambda^{(d)}_{ijk} &=&  \mathcal{R}_{i i^{'}} \mathcal{R}_{j j^{'}} \mathcal{R}_{k k^{'}}  \Lambda^{(d)}_{i^{'} j^{'} k^{'}} , \label{Eq:SymRel_2} \\
\beta_{ijkl} &=&  \mathcal{R}_{i i^{'}} \mathcal{R}_{j j^{'}} \mathcal{R}_{k k^{'}} \mathcal{R}_{l l^{'}}  \beta_{i^{'} j^{'} k^{'} l^{'}}, \label{Eq:SymRel_3} \\
P_{ijkln} &=& |\mathcal{R}|  \mathcal{R}_{i i^{'}} \mathcal{R}_{j j^{'}} \mathcal{R}_{k k^{'}} \mathcal{R}_{l l^{'}} \mathcal{R}_{n n^{'}}  P_{i^{'} j^{'} k^{'} l^{'} n^{'}}. \label{Eq:SymRel_4}
\end{eqnarray}
From these transformation rules, we observe that $\Lambda^{(r)}_{ij}$ and $P_{ijkln}$ are axial tensors, whereas $\Lambda^{(d)}_{ijk}$ and $\beta_{ijkl}$ are polar tensors. 
Importantly, we note that $\Lambda^{(r)}_{ij}= \Lambda^{(d)}_{ijk} = 0$ in systems that are invariant under spatial inversion (i.e., under $\mathcal{R}_{ij}= -\delta_{ij}$).
Thus, as it is established, charge pumping and current-induced torques are absent in homogeneous ferromagnets with inversion symmetry.~\cite{Chernyshov:nature09, Miron:nature10, Miron:n2011, Miron:nm2011, Fang:nn2013, Liu:prl2012, Liu:science2012,  Emori:arxiv2013, Thomas:nnt2013, Garello:nn2013, Haazen:nm2013, Fan:nc2013, Kurebayashi:arxiv2013, Emori:prb2014,Bernevig:prb05, Manchon:prb08, Garate:prb09, Hals:epl10, Pesin:prb2012, Bijl:prb2012, Wang:prl2012, Freimuth:prb2014}  
Only magnetization gradients can mediate a coupling between
the itinerant charge carriers and the magnetization in such systems.  The tensors $\Lambda^{(r)}_{ij}$,  $\Lambda^{(d)}_{ijk}$, $\hat{\Lambda}^{(r)}_{ij}$,  $\hat{\Lambda}^{(d)}_{ijk}$, $\tilde{\Lambda}^{(r)}_{ij}$,  and $\tilde{\Lambda}^{(d)}_{ijk}$
play a role exclusively for crystals described by a non-centrosymmetric point group.  

Next, we apply the general theory developed in this section to two classes of systems:
isotropic systems and asymmetric systems with Rashba and Dresselhaus SOC.
An example of an isotropic system with SOC is the disordered alloy Permalloy, whereas the ferromagnetic semiconductor (Ga,Mn)As is an
example of a system with both large Rashba and Dresselhaus SOC.  

To circumvent the complications introduced by magnetoresistance effects (as mentioned above), we will express the torques in terms of the electric field instead of the charge current density.

\section{Isotropic Systems} \label{Sec:Iso}
Isotropic systems are invariant under any proper or improper rotation.
Thus, spatial inversion is a symmetry operation, which implies that 
\begin{equation}
\Lambda^{(r)}_{ij}= \Lambda^{(d)}_{ijk}= \tilde{\Lambda}^{(r)}_{ij}= \tilde{\Lambda}^{(d)}_{ijk} = \hat{\Lambda}^{(r)}_{ij}= \hat{\Lambda}^{(d)}_{ijk}=  0 . 
\end{equation}
Therefore, the coupled dynamics of the itinerant quasi-particles and the magnetization is governed by the last two terms of the expansions  
in Eq.~\eqref{Eq:ExpLambdaEta} (parameterized by the $\beta$ and $P$ tensors).

\subsection{Invariant tensors of isotropic systems}\label{Sec:InvTenIso}
It is well known that any isotropic tensor of rank two is proportional to the Kronecker delta $\delta_{ij}$, whereas all isotropic third-rank tensors are 
proportional the Levi-Civita tensor $\epsilon_{ijk}$. Furthermore, it has been shown that all isotropic tensors of even rank are linear combinations of products of Kronecker deltas,
whereas isotropic tensors of odd rank are linear combinations of terms formed from products of the Kronecker delta and the Levi-Civita tensor.~\cite{Kearsley:1975}  

Because the $\beta$ and $P$ tensors in Eq.~\eqref{Eq:ExpLambdaEta} are of rank four and five, respectively, we concentrate on these two types of tensors. 
In general, an isotropic tensor $T_{ijkl}^{(4)}$ of rank four has three independent tensor coefficients  and can be written as~\cite{Kearsley:1975} 
\begin{equation}
T_{ijkl}^{(4)} = T_1^{(4)} \delta_{ij}\delta_{kl} + T_2^{(4)} \delta_{ik}\delta_{jl} + T_3^{(4)} \delta_{il}\delta_{jk} \label{Eq:T4}. 
\end{equation}
An isotropic tensor $T_{ijkln}^{(5)}$ of rank five has six independent coefficients ~\cite{Kearsley:1975}  
\begin{eqnarray}
T_{ijkln}^{(5)} &=& T_1^{(5)}\epsilon_{ijk} \delta_{ln} + T_2^{(5)}\epsilon_{ijl} \delta_{kn}  + T_3^{(5)}\epsilon_{ijn} \delta_{kl} +    \nonumber \\
& & T_4^{(5)}\epsilon_{ikl} \delta_{jn} + T_5^{(5)}\epsilon_{ikn} \delta_{lj} + T_6^{(5)}\epsilon_{iln} \delta_{jk} \label{Eq:T5}.  
\end{eqnarray}
Consequently, $\hat{\beta}_{ijkl}$ and $\tilde{\beta}_{ijkl}$ have tensorial forms given by Eq.~\eqref{Eq:T4}, whereas the tensorial forms of $\hat{P}_{ijkln}$ and $\tilde{P}_{ijkln}$ are determined by Eq.~\eqref{Eq:T5}. 

We will later on apply the following useful identity ~\cite{Kearsley:1975}
\begin{equation}
\epsilon_{ijk}\delta_{mp} - \epsilon_{jkm}\delta_{ip} + \epsilon_{kmi}\delta_{jp} - \epsilon_{mij}\delta_{kp}  = 0, \label{Eq:Capelli}
\end{equation}
which can be used to decompose an isotropic tensor of rank five into a linear combination of other terms. 

\subsection{Expressions for torque and charge pumping}\label{Sec:IsoTorquePumping}
In the following discussion, it is convenient to separate the charge pumping and torque terms into dissipative and reactive contributions.
The reactive terms arise from $\hat{P}_{ijkln}$ and $\tilde{P}_{ijkln}$, whereas the dissipative terms are determined by the tensors $\hat{\beta}_{ijkl}$ and $\tilde{\beta}_{ijkl}$.

Using Eq.~\eqref{Eq:T4}, we find the dissipative current-induced field $\mathbf{H}_c^{(d)}$ and charge pumping $\boldsymbol{\mathcal{\tilde{J}}}^{(d)}$ 
\begin{eqnarray} 
\mathcal{\tilde{J}}_i^{(d)} &=& \tilde{\beta}_1 \dot{\mathbf{m}}\cdot\partial_i\mathbf{m} +  \tilde{\beta}_2\dot{\mathbf{m}}\cdot\boldsymbol{\nabla} m_i +   \tilde{\beta}_3  \dot{m_i} \boldsymbol{\nabla}\cdot \mathbf{m}, \label{Eq:Jdiso} \\
H_{c,i}^{(d)} &=& \frac{1}{M_s} \left[ \hat{\beta}_1\mathbf{E}\cdot\boldsymbol{\nabla} m_i + \hat{\beta}_2 \mathbf{E}\cdot \partial_i \mathbf{m} + \hat{\beta}_3 E_i \boldsymbol{\nabla}\cdot\mathbf{m} \right] . \nonumber
\end{eqnarray}
The reactive field $\mathbf{H}_c^{(r)}$ and charge pumping $\boldsymbol{\mathcal{\tilde{J}}}^{(r)}$ each have six independent terms.
However, because of the normalization condition $\mathbf{m}\cdot\mathbf{m} = 1$, one of these terms vanishes and the resulting expressions for the reactive field and charge pumping become 
\begin{eqnarray} 
\mathcal{\tilde{J}}_i^{(r)} &=& \tilde{P}_1 \mathbf{m}\cdot \left(\dot{\mathbf{m}}\times \partial_i \mathbf{m} \right) + \tilde{P}_2 \left(\dot{\mathbf{m}}\times  \mathbf{m} \right)_i \left( \boldsymbol{\nabla}\cdot\mathbf{m} \right)  \nonumber \\
& & +\  \tilde{P}_3  \left(\dot{\mathbf{m}}\times \left(\mathbf{m}\cdot\boldsymbol{\nabla}\right) \mathbf{m}  \right)_i +  \tilde{P}_4 \mathbf{m}\cdot \left( \dot{\mathbf{m}}\times\boldsymbol{\nabla} \right)  m_i  \nonumber  \\
& &  -\ \tilde{P}_5 m_i \dot{\mathbf{m}}\cdot \left( \boldsymbol{\nabla}\times\mathbf{m} \right) , \label{Eq:Jriso} \\ 
H_{c,i}^{(r)} &=& \frac{1}{M_s} [ \hat{P}_1\left( \mathbf{m}\times \left(\mathbf{E}\cdot\boldsymbol{\nabla} \right)\mathbf{m}\right)_i +  \hat{P}_2\left( \boldsymbol{\nabla}\cdot\mathbf{m} \right)  \left(\mathbf{E} \times\mathbf{m} \right)_i  \nonumber \\
& & +\ \hat{P}_3\left( \mathbf{E}\times \left(\mathbf{m}\cdot\boldsymbol{\nabla}\right) \mathbf{m}  \right)_i + \hat{P}_4 \mathbf{E} \cdot \left( \mathbf{m} \times \boldsymbol{\nabla}  \right)_i \mathbf{m}  \nonumber \\
& & +\ \hat{P}_5 \left(\mathbf{E}\cdot\mathbf{m}\right) \left( \boldsymbol{\nabla}\times \mathbf{m} \right)_i ] . \label{Eq:Hriso}
\end{eqnarray}
The Onsager reciprocal relations \eqref{Eq:Onsager_3} - \eqref{Eq:Onsager_4}  imply that the phenomenological parameters in Eq.~\eqref{Eq:Jdiso} - \eqref{Eq:Hriso} are related via the equations
\begin{alignat}{2}
\tilde{\beta}_i &= \hat{\beta}_{i},  \ \ \ \ \ \  & i &\in \left\{ 1, 2, 3 \right\} , \\
\tilde{P}_i &=  \hat{P}_i ,   \ \ \ \ \ \  & i &\in \left\{ 1, ..., 5 \right\} .
\end{alignat}

Now, we will demonstrate that the conventional LLGS equation \eqref{Eq:LLG} and charge pumping \eqref{Eq:ChargePumpingSTT} represent a particular limiting case for isotropic systems.

\subsection{Non-relativistic limit: conventional LLGS} 
To illustrate how the system is affected by the SOC, we consider the SOC Hamiltonian
\begin{equation}
H_{\rm so} = \frac{\hbar}{4m^2c^2}\left( \boldsymbol{\nabla} V \times\mathbf{\hat{p}} \right)\cdot\boldsymbol{\hat{\sigma}} . \label{Eq:SOC}
\end{equation} 
Here, $V(\mathbf{r})$ is the crystal potential, $\mathbf{\hat{p}}$ is the momentum operator, $\boldsymbol{\hat{\sigma}}$ is the vector of Pauli matrices, $m$ is the electron mass, and $c$ is the speed of light.
Importantly, $H_{\rm so}$ vanishes in the non-relativistic limit $c\rightarrow \infty$. 
The Hamiltonian \eqref{Eq:SOC} links the spin space to the symmetry of the underlying crystal structure. For an isotropic system, Eq.~\eqref{Eq:SOC} is (after disorder averaging) invariant under any proper or improper 
rotation $\boldsymbol{\mathcal{R}}$ that acts \emph{simultaneously on the spin space and the real space}, i.e., acting on both the spatial vectors (such as $\boldsymbol{\nabla}V$ and $\mathbf{\hat{p}}$) and the spin vector $\boldsymbol{\hat{\sigma}}$. 
However, without SOC, the spin space and real space are no longer coupled and can therefore be separately rotated. This decoupling increases the symmetry of the system:  an isotropic system without SOC is invariant under 
\emph{separate rotations of the spin space and the real space}.

The expressions in Sec.~\ref{Sec:IsoTorquePumping} are invariant under simultaneous rotations of spatial vectors and spin vectors, i.e., they apply to isotropic systems with SOC.
If we also require that the expressions are invariant under separate rotations of the spin space and real space, we find that only  the dissipative contributions proportional to $\hat{\beta}_1, \tilde{\beta}_1$ and
the reactive contributions proportional to $\hat{P}_1, \tilde{P}_1$ satisfy the correct symmetry. 
Without SOC, the conductivity tensor becomes $\sigma_{ij}= \sigma \delta_{ij}$ and $\boldsymbol{\hat{\eta}}=  M_s \sigma \boldsymbol{\eta}$.
Thus, for an isotropic system without SOC, the charge pumping and torque are given by
\begin{eqnarray}
\mathcal{\tilde{J}}_i &= & \tilde{P}_1 \mathbf{m}\cdot \left(\dot{\mathbf{m}}\times \partial_i \mathbf{m} \right) +  \tilde{\beta}_1 \dot{\mathbf{m}}\cdot\partial_i\mathbf{m} ,\label{Eq:JisoNonRel} \\
\boldsymbol{\tau} &=& \frac{\gamma \hat{P}_1}{M_s \sigma } \left(\boldsymbol{\mathcal{J}}\cdot\boldsymbol{\nabla} \right)\mathbf{m} - \frac{\gamma \hat{\beta}_1}{M_s \sigma} \mathbf{m}\times \left( \boldsymbol{\mathcal{J}}\cdot\boldsymbol{\nabla}\right) \mathbf{m} . \label{Eq:tauIsoNonRel}
\end{eqnarray}     

Eq.~\eqref{Eq:JisoNonRel} is identical to the charge pumping in Eq.~\eqref{Eq:ChargePumpingSTT}, whereas Eq.~\eqref{Eq:tauIsoNonRel} reduces to the STT in Eq.~\eqref{Eq:STT} when $\gamma \hat{P}_1\boldsymbol{\mathcal{J}}/M_s\sigma = - \mathbf{v}_s $ and 
$ \gamma \hat{\beta}_1\boldsymbol{\mathcal{J}}/M_s\sigma= - \beta \mathbf{v}_s$. 

From the above analysis, we conclude that the conventional LLGS equation  \eqref{Eq:LLG} and charge pumping  \eqref{Eq:ChargePumpingSTT}  are obtained as a particular limiting case for our phenomenology:
\emph{the non-relativistic limit of an isotropic system}. 

\subsection{The relativistic corrections} 
Our phenomenology shows that the SOC introduces two new dissipative terms and four new reactive terms in isotropic systems such as Permalloy.
Thus far, none of these terms have been inferred from experiments. 
However, if these terms are comparable in magnitude to the conventional, non-relativistic  terms, they have important consequences for the current-driven dynamics of domain walls.
Because $\boldsymbol{\nabla}\cdot\mathbf{m}=0$ ($\boldsymbol{\nabla}\cdot\mathbf{m}\neq 0$) for Bloch (Neel) walls, the terms proportional to $\hat{P}_2$ and $\hat{\beta}_3$ only contribute to the dynamics of Neel walls,
which is also the case for the other new terms; the terms contribute differently for different types of domain walls. Therefore, even in isotropic systems, Bloch and Neel walls are expected to exhibit different current-driven drift velocities.

A way to probe some of the relativistic corrections is to investigate the current-driven domain wall drift velocity for electric fields applied parallel to the wall. In this case, only the relativistic corrections to the reactive
STT is able to drive domain wall motion.  
Let us consider Bloch and Neel walls of the forms $\mathbf{m} (z,t) = [s_1 {\rm sech} (\tilde{z}), s_2 {\rm tanh} (\tilde{z}) , 0]$ and $\mathbf{m} (z,t) = [0, s_1 {\rm sech}(\tilde{z}), s_2 {\rm tanh} (\tilde{z})]$, respectively. Here, $\tilde{z} = (z - r_{dw})/\lambda$ 
where $r_{dw}$ is the domain wall position along the $z$ axis, $\lambda$ is the domain wall width, and $s_1\in \{ -1, 1 \}$, $s_2\in \{ -1, 1 \}$. 
Using the collective coordinate description outlined in Ref.~\onlinecite{Tretiakov:prl2008}, we find for an electric field $\mathbf{E}= [E_x, E_y, 0]$ the following stationary drift velocities for Bloch and Neel walls
\begin{eqnarray}
{\rm Bloch:}\  \  \dot{r}_{dw} &=& -s_{1} \frac{\pi P_4}{4 \alpha} E_y, \label{Eq:drdt1}  \\  
{\rm Neel:} \ \  \dot{r}_{dw} &=& -s_{1} \frac{\pi P_2}{4 \alpha} E_x . \label{Eq:drdt2}
\end{eqnarray} 
In a real experiment, it is difficult to produce an electric field that is perfectly parallel to the wall. Therefore, in practice, there will most likely also be a small $E_z$ component. This $E_z$ component can induce a drift velocity via the conventional dissipative STT, i.e.,  $\dot{r}_{dw} \sim \beta/\alpha$. However, Eqs.~\eqref{Eq:drdt1}-\eqref{Eq:drdt2} implies that a reversed domain wall, $\mathbf{m}\rightarrow -\mathbf{m}$, moves in the opposite direction. This differs from the drift velocity induced by the dissipative STT, which is unchanged under $\mathbf{m}\rightarrow -\mathbf{m}$. Thus, if measurements of up-down and down-up domain walls yield opposite drift velocities, this is a clear indicator that the relativistic reactive STT is the driving mechanism behind the current-induced velocity.   

Some of the relativistic corrections in Eqs.~\eqref{Eq:Jdiso}-\eqref{Eq:Hriso} can be attributed to the spin Hall effect (SHE). Ref.~\onlinecite{Manchon:apl2011} showed that the SHE produces 
reactive and dissipative torques of the form
\begin{eqnarray}
\boldsymbol{\tau}_{she}^{(d)} &=& -\beta \alpha_{H} (\mathbf{E}\times\mathbf{m})\cdot\boldsymbol{\nabla}\mathbf{m} , \\
\boldsymbol{\tau}_{she}^{(r)} &=&   \alpha_{H}\mathbf{m}\times [ (\mathbf{E}\times\mathbf{m})\cdot\boldsymbol{\nabla}\mathbf{m} ] .
\end{eqnarray}
Here, $\alpha_H$ is the Hall angle. 
Using the identity in Eq.~\eqref{Eq:Capelli}, the dissipative SHE torque can be decomposed into
\begin{eqnarray}
\tau_{she, i}^{(d)} &=& -\beta \alpha_{H} [ \mathbf{E}\cdot (\mathbf{m}\times \boldsymbol{\nabla} )_i \mathbf{m} - (\mathbf{m}\times\mathbf{E})_i (\boldsymbol{\nabla}\cdot\mathbf{m})  ] . \nonumber
\end{eqnarray} 
The first term is of the same form as the $\hat{\beta}_2$ torque in Eq.~\eqref{Eq:Jdiso}, while the second term has the form of the $\hat{\beta}_3$ torque. Taking the cross product with $\mathbf{m}$, one finds from the above expression
that reactive torque $\boldsymbol{\tau}_{she}^{(r)}$ can be written as a linear combination of the $\hat{P}_2$ and $\hat{P}_4$ terms in Eq.~\eqref{Eq:Hriso}.  
The reciprocal processes of these SHE torques were calculated microscopically in Ref.~\onlinecite{Shibata:prl2009}. 
  
We believe an important task for future experiments is to investigate the magnitude of the relativistic torques in isotropic ferromagnets such as Permalloy. Additionally, improved microscopic theories are required to
gain a deeper understanding of the relativistic corrections to the current-driven magnetization dynamics and to determine the magnitude of the new terms.

\section{Rashba and Dresselhaus SOC} \label{Sec:C2v}
\begin{figure}[t] 
\centering 
\includegraphics[scale=1.0]{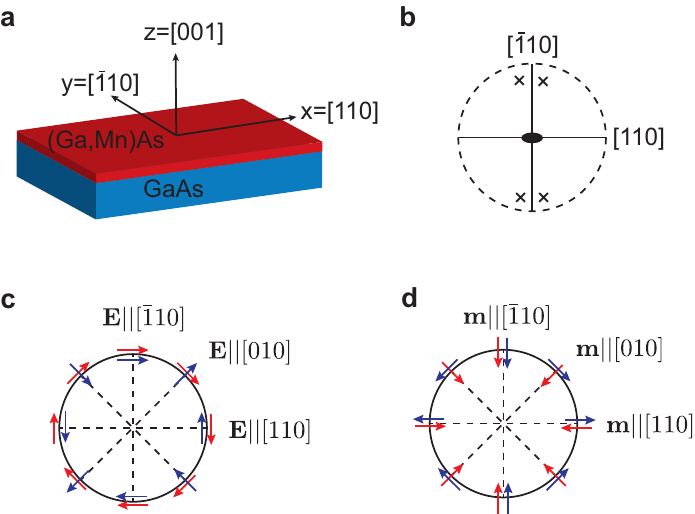}  
\caption{(Color online). (a) A (Ga,Mn)As epilayer grown on a GaAs substrate along the crystallographic direction $[001]$. (b) Stereographic projection of the point group $C_{2v}$, which illustrates the symmetry of strained (Ga,Mn)As.
(c) Vector plot of the reactive SOT fields induced by Rashba (red arrows) and Dresselhaus (blue arrows) SOC for different directions of the applied electric field. (d) Vector plot of the in-plane current pumping due to magnetization precession for different directions of the equilibrium magnetization. The charge pumping via Rashba (Dresselhaus) SOC is illustrated by red (blue) arrows. The reactive and dissipative contributions to the pumped in-plane current have identical symmetries.  }
\label{Fig2} 
\end{figure}

Next, we consider asymmetric ferromagnets with Rashba and Dresselhaus SOC. These systems are described by the crystallographic point group $C_{2v}$.
An important example of such a system is strained (Ga,Mn)As grown on GaAs along the crystallographic axis $[001]$ (Fig.~\ref{Fig2}a-b). In this case,  
the two-fold symmetry axis is perpendicular to the epilayer (i.e., along $[001]$), and the two reflection planes are defined by the axes $[110]$ and $[\bar{1}10]$. 
Herein, we define the reference frame such that the two-fold axis is along $z$, whereas the $x$ and $y$ axes define the two reflection planes.  
For (Ga,Mn)As, this reference frame is related to the crystallographic axes via $\hat{\mathbf{x}}= [110]$, $\hat{\mathbf{y}}= [\bar{1}10]$, and $\hat{\mathbf{z}}= [001]$. 
We consider ferromagnets with a uniform magnetization and model the coupled dynamics 
of the magnetization and itinerant charge carriers using the four second- and third-rank tensors
$\hat{\Lambda}^{(r)}$, $\tilde{\Lambda}^{(r)}$, $\hat{\Lambda}^{(d)}$, and $\tilde{\Lambda}^{(d)}$ in Eq.~\eqref{Eq:ExpLambdaEta}.
Note that these terms originate solely from the SOC and vanish in systems with spatial inversion symmetry. Thus, $\hat{\Lambda}^{(r)}$ and $\hat{\Lambda}^{(d)}$ parameterize an SOT, whereas $\tilde{\Lambda}^{(r)}$ and   
$\tilde{\Lambda}^{(d)}$ describe charge pumping caused by a direct conversion of zero wavevector magnons into an electric current via the SOC, which we previously termed \emph{magnonic charge pumping}.~\cite{Chiara:nn2015} 

\subsection{Invariant tensors of the $C_{2v}$ point group} 
The invariant axial second-rank tensors $T_{ij}^{(2)}$ of the point group $C_{2v}$ are described by two independent tensor coefficients.
For the reference frame defined above, the two non-vanishing tensor coefficients are~\cite{Birss:book}
\begin{equation}
T_{xy}^{(2)} , \  \  \ T_{yx}^{(2)}  . \label{Eq:T2C2v}
\end{equation}   
Polar third-rank tensors $T_{ijk}^{(3)}$ that are invariant under $C_{2v}$ are described by the following seven independent tensor coefficients~\cite{Birss:book}
\begin{equation}
T_{zzz}^{(3)} , \  \  \ T_{xxz}^{(3)}  , \ \ \ T_{zxx}^{(3)}  , \  \  \ T_{xzx}^{(3)}  , \ \ \   T_{yyz}^{(3)}  , \  \  \ T_{zyy}^{(3)}  , \ \ \  T_{yzy}^{(3)}  . \label{Eq:T3C2v}
\end{equation}
Thus,  charge pumping and SOT are described by nine independent tensor coefficients. 
The tensors  $\hat{\Lambda}^{(r)}$ and $\tilde{\Lambda}^{(r)}$ are characterized by the two elements in Eq.~\eqref{Eq:T2C2v}, whereas $\hat{\Lambda}^{(d)}$ and $\tilde{\Lambda}^{(d)}$ are 
determined by the seven coefficients of Eq.~\eqref{Eq:T3C2v}. 

\subsection{Expressions for torque and charge pumping}   
We separate the various terms into reactive and dissipative contributions. 
The reactive charge pumping and SOT field are determined by the second-rank tensors  $\tilde{\Lambda}^{(r)}$ and $\hat{\Lambda}^{(r)}$, respectively, which lead to the expressions  
\begin{eqnarray}
\mathcal{\tilde{J}}_i^{(r)} &=& \delta_{ix} \tilde{\Lambda}_{xy}^{(r)}\dot{m}_y + \delta_{iy} \tilde{\Lambda}_{yx}^{(r)}\dot{m}_x,\label{Eq:JrC2v} \\
M_s H_{c,i}^{(r)} &=& \delta_{ix} \hat{\Lambda}_{xy}^{(r)} E_y + \delta_{iy} \hat{\Lambda}_{yx}^{(r)} E_x  . \label{Eq:HrC2v}
\end{eqnarray}
The dissipative charge pumping and SOT are given by the third-rank tensors $\tilde{\Lambda}^{(d)}$ and $\hat{\Lambda}^{(d)}$, respectively.
From Eq.~\eqref{Eq:T3C2v}, we find the pumped current and the effective field:
\begin{eqnarray}
\mathcal{\tilde{J}}_i^{(d)} &=& \delta_{ix}\left[ \tilde{\Lambda}_{xxz}^{(d)}\dot{m}_x m_z + \tilde{\Lambda}_{xzx}^{(d)}\dot{m}_z m_x  \right] + \label{Eq:JdC2v}  \\
& & \delta_{iy}\left[ \tilde{\Lambda}_{yyz}^{(d)}\dot{m}_y m_z + \tilde{\Lambda}_{yzy}^{(d)}\dot{m}_z m_y  \right]  + \nonumber  \\
& & \delta_{iz}\left[ \tilde{\Lambda}_{zzz}^{(d)}\dot{m}_z m_z + \tilde{\Lambda}_{zxx}^{(d)}\dot{m}_x m_x + \tilde{\Lambda}_{zyy}^{(d)}\dot{m}_y m_y  \right]  , \nonumber \\
M_s H_{c,i}^{(d)} &=& \delta_{ix}\left[ \hat{\Lambda}_{xxz}^{(d)} E_x m_z + \hat{\Lambda}_{xzx}^{(d)} E_z m_x  \right] + \label{Eq:HdC2v}  \\
& & \delta_{iy}\left[ \hat{\Lambda}_{yyz}^{(d)}E_y m_z + \hat{\Lambda}_{yzy}^{(d)}E_z m_y  \right]  + \nonumber  \\
& & \delta_{iz}\left[ \hat{\Lambda}_{zzz}^{(d)}E_z m_z + \hat{\Lambda}_{zxx}^{(d)}E_x m_x + \hat{\Lambda}_{zyy}^{(d)}E_y m_y  \right]  . \nonumber
\end{eqnarray}

The reciprocal relationships between the phenomenological pumping and torque parameters are determined by Eqs.~\eqref{Eq:Onsager_1} - \eqref{Eq:Onsager_2}.

As an application of  Eqs.~\eqref{Eq:JrC2v} - \eqref{Eq:HdC2v}, we focus on the specific system of (Ga,Mn)As, for which the SOT and charge pumping have been
experimentally observed.~\cite{Chiara:nn2015}

\subsection{Example: (Ga,Mn)As epilayers} 
We consider a strained (Ga,Mn)As epilayer grown on a GaAs $[001]$ substrate. At equilibrium, the magnetization is assumed to align along an in-plane direction (i.e., perpendicular to $[001]$).
In addition, we only consider in-plane electric fields and pumped charge currents: 
\begin{equation}
\mathbf{m} = \left[ m_x\  m_y \  0 \right] , \ \ 
\mathbf{E} =   \left[  E_x\  E_y\  0 \right]  , \ \ 
\boldsymbol{\mathcal{\tilde{J}}} =   \left[  \tilde{\mathcal{J}}_x\  \tilde{\mathcal{J}}_y\  0 \right] .   \nonumber
\end{equation}
We parameterize the direction of $\mathbf{m}$ and $\mathbf{E}$ by the angles $\phi_m$ and $\phi_e$, respectively, such that
$E_x= E_0\cos (\phi_e)$, $E_y= E_0\sin (\phi_e)$, $m_x= \cos (\phi_m)$, and $m_y= \sin (\phi_m)$.
These angles are measured from the $x$ axis, which is parallel to the crystallographic direction $[110]$. 
For later use, we also introduce the angle $\phi_{me}$, which represents the relative angle between the magnetization and the applied electric field
$\phi_{me}\equiv \phi_m - \phi_e$.  

Because the magnetization and applied electric field do not have a $z$ component, several terms in Eqs.~\eqref{Eq:JrC2v} - \eqref{Eq:HdC2v} vanish.
The reactive charge pumping and current-induced field remain characterized by the same two tensor coefficients, as given above.
However, the dissipative contributions become considerably simplified. 
The dissipative (in-plane) charge pumping is governed only by the two terms that are proportional to $\tilde{\Lambda}_{xzx}^{(d)}$ and $\tilde{\Lambda}_{yzy}^{(d)}$, whereas
the dissipative SOT field is given by  $\hat{\Lambda}_{zxx}^{(d)}$ and $\hat{\Lambda}_{zyy}^{(d)}$. Because the pumping and torque parameters are related
via the reciprocal relations, we conclude that only four independent parameters are required to model this system. It is convenient to decompose the four terms into
Rashba-like and Dresselhaus-like contributions by introducing the parameters  
\begin{alignat}{2}
\tilde{\Lambda}_R^{(r)} &= ( \tilde{\Lambda}_{yx}^{(r)} - \tilde{\Lambda}_{xy}^{(r)} )/2 , \ \  &\tilde{\Lambda}_D^{(r)} &= ( \tilde{\Lambda}_{xy}^{(r)} + \tilde{\Lambda}_{yx}^{(r)} )/2 ,\nonumber \\ 
\tilde{\Lambda}_R^{(d)} &= ( \tilde{\Lambda}_{xzx}^{(d)} + \tilde{\Lambda}_{yzy}^{(d)} )/2 , \ \  &\tilde{\Lambda}_D^{(d)} &= ( \tilde{\Lambda}_{xzx}^{(d)} - \tilde{\Lambda}_{yzy}^{(d)} )/2,\nonumber \\
\hat{\Lambda}_R^{(r)} &= ( \hat{\Lambda}_{xy}^{(r)} - \hat{\Lambda}_{yx}^{(r)} )/2 , \ \  & \hat{\Lambda}_D^{(r)} &= ( \hat{\Lambda}_{xy}^{(r)} + \hat{\Lambda}_{yx}^{(r)} )/2 , \nonumber \\  
\hat{\Lambda}_R^{(d)} &= ( \hat{\Lambda}_{zxx}^{(d)} + \hat{\Lambda}_{zyy}^{(d)} )/2 , \ \  & \hat{\Lambda}_D^{(d)} &= ( \hat{\Lambda}_{zxx}^{(d)} - \hat{\Lambda}_{zyy}^{(d)} )/2. \label{Eq:Lambda_RD}
\end{alignat}
Here, $\hat{\Lambda}_R^{(r)}$ ($\tilde{\Lambda}_R^{(r)}$) and $\hat{\Lambda}_D^{(r)}$ ($\tilde{\Lambda}_D^{(r)}$) parameterize the reactive SOT (pumping) caused by Rashba and Dresselhaus SOC, respectively, whereas 
$\hat{\Lambda}_R^{(d)}$ ($\tilde{\Lambda}_R^{(d)}$) and $\hat{\Lambda}_D^{(d)}$ ($\tilde{\Lambda}_D^{(d)}$) represent the corresponding dissipative processes. By describing Eqs.~\eqref{Eq:JrC2v} - \eqref{Eq:HdC2v} in terms of the
parameters in  Eq.~\eqref{Eq:Lambda_RD}, we obtain the expressions 
\begin{widetext}
\begin{eqnarray}
\begin{pmatrix}
\tilde{\mathcal{J}}_x \\
\tilde{\mathcal{J}}_y
\end{pmatrix}
&=&
\tilde{\Lambda}_{R}^{(r)}
\begin{pmatrix}
0 & -1  \\
1 & 0
\end{pmatrix}
\begin{pmatrix}
\dot{m}_x \\
\dot{m}_y
\end{pmatrix}
+
\tilde{\Lambda}_{D}^{(r)}
\begin{pmatrix}
0 & 1  \\
1 & 0
\end{pmatrix}
\begin{pmatrix}
\dot{m}_x \\
\dot{m}_y
\end{pmatrix}
+
\tilde{\Lambda}_{R}^{(d)}
\dot{m}_z
\begin{pmatrix}
1 & 0  \\
0 & 1
\end{pmatrix}
\begin{pmatrix}
m_{ x} \\
m_{ y}
\end{pmatrix}
+
\tilde{\Lambda}_{D}^{(d)}
\dot{m}_z
\begin{pmatrix}
1 & 0  \\
0 & -1
\end{pmatrix}
\begin{pmatrix}
m_{ x} \\
m_{y}
\end{pmatrix}  , \label{Eq:PumpIPCv2} \\
\begin{pmatrix}
H_{c,x} \\
H_{c,y}
\end{pmatrix}
&=&
\frac{\hat{\Lambda}_R^{(r)}}{M_s}
\begin{pmatrix}
0 & 1  \\
-1 & 0
\end{pmatrix}
\begin{pmatrix}
E_x \\
E_y
\end{pmatrix}
+
\frac{\hat{\Lambda}_D^{(r)}}{M_s}
\begin{pmatrix}
0 & 1  \\
1 & 0
\end{pmatrix}
\begin{pmatrix}
E_x \\
E_y
\end{pmatrix}
 , \label{Eq:FieldIPCv2} \\
H_{c,z} &=& \frac{\hat{\Lambda}_R^{(d)}}{M_s} ( E_x m_{x}  + E_y m_{y}  ) + \frac{\hat{\Lambda}_D^{(d)}}{M_s} ( E_x m_{x}  - E_y m_{y}  ) . \label{Eq:FieldOPCv2} 
\end{eqnarray}
\end{widetext}

The charge pumping of Eq.~\eqref{Eq:PumpIPCv2} was recently investigated and measured in Ref.~\onlinecite{Chiara:nn2015}. 
The reactive and dissipative charge pumping with Rashba symmetry agree with the results derived microscopically for a Rashba model in Refs.~\onlinecite{Kim:prl2012, Tatara:prb2013}.
The reactive parts of the SOT in Eq.~\eqref{Eq:FieldIPCv2} have been derived microscopically in several works,\cite{Bernevig:prb05, Manchon:prb08, Obata:prb08, Garate:prb09} while the dissipative SOT \eqref{Eq:FieldOPCv2} agrees with the results derived in  Refs. ~\onlinecite{Pesin:prb2012, Wang:prl2012, Kurebayashi:arxiv2013}.

The charge-pumping and reactive SOT-field symmetry of Eqs.~\eqref{Eq:PumpIPCv2} - \eqref{Eq:FieldOPCv2} can be represented in a vector plot.
Fig.~\ref{Fig2}c-d shows the direction of the reactive and dissipative pumped charge current for different directions of the equilibrium magnetization and the direction of the reactive 
SOT field for different directions of the applied electric field. Recent experiments have verified that the SOT and charge pumping in (Ga,Mn)As follow the forms of Eqs.~\eqref{Eq:PumpIPCv2} - \eqref{Eq:FieldOPCv2}, as implied by our phenomenology.~\cite{Chernyshov:nature09, Fang:nn2013, Kurebayashi:arxiv2013, Chiara:nn2015}

Because both reactive and dissipative processes contribute to the in-plane charge pumping, these processes can be difficult to disentangle from a measurement of the charge pumping.
However, for the SOT field, only the reactive processes control the in-plane components of the field, whereas the out-of-plane component $H_{c,z}$ is 
governed by the dissipative effects. 

The magnitude of $H_{c,z}$ can be written as a function of two angles, $\phi_{me}$ and  $\phi_e$.
From Eq~\eqref{Eq:FieldOPCv2}, we find 
\begin{equation}
H_{c,z}= \frac{\hat{\Lambda}_R^{(d)}}{M_s} E_0 \cos (\phi_{me}) + \frac{\hat{\Lambda}_D^{(d)}}{M_s} E_0 \cos (\phi_{me} + 2\phi_e) .
\end{equation}
The $\phi_{me}$ dependence of $H_{c,z}$ is illustrated in Tab.~\ref{tab:2} for four different directions of the applied electric field and is consistent with
the out-of-plane field observed in Ref. ~\onlinecite{Kurebayashi:arxiv2013}.

\begin{table}
\centering
\caption{ Angle $\phi_{me}$ dependence of the Rashba and Dresselhaus contributions to the SOT field component $H_{c,z}$. }
\begin{tabular}{ l c c c }
\hline\hline \\
& \ \ \ \   $\phi_e$\ \  & \ \ \ \  Rashba SOC\ \ \ \   &  \ \ \ \  Dresselhaus SOC \ \ \ \  \\
 \hline \\
$\mathbf{E} || [100]$ & $-\frac{\pi}{4}$ & $\cos (\phi_{me})$ & $\sin (\phi_{me})$ \\ 
$\mathbf{E} || [010]$ & $\frac{\pi}{4}$ & $\cos (\phi_{me})$ & $-\sin (\phi_{me})$  \\
$\mathbf{E} || [110]$ & 0 &  $\cos (\phi_{me})$ & $\cos (\phi_{me})$  \\
$\mathbf{E} || [1\bar{1}0]$ & $-\frac{\pi}{2}$ & $\cos (\phi_{me})$ & $-\cos (\phi_{me})$  \\
 \hline \hline	 
\end{tabular}
\label{tab:2}
\end{table}

The reactive and dissipative charge pumping caused by Rashba (Dresselhaus) SOC are related via the Onsager reciprocal relations in Eqs.~\eqref{Eq:Onsager_1} - \eqref{Eq:Onsager_2}: 
\begin{alignat}{2}
\tilde{\Lambda}_R^{(r)} &=  - \hat{\Lambda}_R^{(r)} , \ \ \ \    & \tilde{\Lambda}_D^{(r)} &=  -   \hat{\Lambda}_D^{(r)} , \label{Eq:ResRelRD1}  \\
\tilde{\Lambda}_R^{(d)} &=    \hat{\Lambda}_R^{(d)} , \ \ \ \   & \tilde{\Lambda}_D^{(d)} &=    \hat{\Lambda}_D^{(d)} . \label{Eq:ResRelRD2}
\end{alignat}
Thus, by measuring one of these two processes (e.g., the SOT), one also automatically gains knowledge regarding the strength of the reciprocal phenomenon (e.g., the charge pumping).
This reciprocal relationship between magnonic charge pumping and SOT was recently demonstrated experimentally in (Ga,Mn)As.\cite{Chiara:nn2015}

\section{Summary}\label{Sec:Summary}
Several experiments have revealed an extreme complexity in the coupled dynamics between itinerant charge carriers and magnetization in systems with strong SOC. 
The theoretical formalism in this paper outlines a systematic scheme for deriving all possible torque and charge pumping terms that 
satisfy the correct symmetry implied by Neumann's principle and relates these terms via the Onsager reciprocal relations.
We hope that the present formalism will play an important role in the exploration of these systems and will be a useful tool 
for understanding and directing future experiments on ferromagnets with strong SOC.

We apply the formalism to two important classes of systems: isotropic ferromagnets with non-uniform magnetization and homogenous ferromagnets described by
the crystallographic point group $C_{2v}$. In isotropic systems, the phenomenology reduces to the conventional LLGS equation and spin-motive force in the non-relativistic limit.
However, we show that the SOC in these systems introduces six new reactive and dissipative terms.
Four of these new terms can be attributed to the SHE, whereas two of the reactive terms have not been reported before.
Importantly, the new terms lead to different dynamics for Bloch and Neel domain walls, i.e.,
the dynamics are dependent on the structure of the magnetic texture. 
For asymmetric ferromagnets described by the point group $C_{2v}$, we concentrate on strained (Ga,Mn)As. 
We show that the SOTs and charge pumping derived from our phenomenology are consistent with recent experimental observations in (Ga,Mn)As epilayers.


\begin{thebibliography}{99}
\bibitem{comment1} By intrinsic SOC, we refer to the SOC associated with a perfect crystalline lattice potential. Impurity scattering potentials are responsible for the extrinsic SOC.
\end{thebibliography}
\end{document}